# Deep learning for improved global precipitation in numerical weather prediction systems

Manmeet Singh[a,b,c*], Bipin Kumar[a], , Suryachandra Rao[a], Sukhpal Singh Gill[a], Rajib Chattopadhyay[a], Ravi S Nanjundiah[a], Dev Niyogi[b]

[a]Indian Institute of Tropical Meteorology, Ministry of Earth Sciences, India
[b]Jackson School of Geosciences, University of Texas at Austin, USA
[c]IDP in Climate Studies, Indian Institute of Technology Bombay, India
[d]School of Electrical Engineering and Computer Science, Queen Mary University of London, UK

* manmeet.singh@utexas.edu

**Abstract.** The formation of precipitation in state-of-the-art weather and climate models is an important process. The understanding of its relationship with other variables can lead to endless benefits. Various factors play a crucial role in the formation of rainfall, and those physical processes are leading to significant biases in the operational weather forecasts. We use the UNET architecture of a deep convolutional neural network with residual learning as a proof of concept to learn global data-driven models of precipitation. TThe results are compared with a operational dynamical model. The theoretical deep learning-based model shows doubling of the grid point, as well as area averaged skill measured in Pearson correlation coefficients relative to operational system. This study is a proof-of-concept showing that residual learning-based UNET can unravel physical relationships to target precipitation, which can be used in the dynamical operational models towards improved precipitation forecasts.

## 1. Introduction

The estimation of precipitation in weather and climate models is a challenging problem. "The real holes in climate science", identified precipitation forecasts in the state-of-the-art dynamical models as one of the four major problems impeding the progress in weather and climate science [1]. According to the Intergovernmental Panel on Climate Change Assessment Report 5 (IPCC AR5), the climate models diverge significantly on the rainfall and snow patterns. More precisely, there are issues with the winter precipitation which is used to replenish the potable water supply. These issues were the reason the IPCC AR5 was unable to conclusively reach a consensus on the winter precipitation. A significant problem that has been identified and which is stymying the predictions of precipitation in the models is vertical convection, particularly in the tropical regions. [2]

review past literature on the subject and show that the general circulation models have biases in the radiative fluxes which are in turn related to the errors in cloud cover. These errors in the parameterizations have been attributed to spoiling precipitation forecasts in the state-of-the-art models due to their non-exact representations. Deep learning can address this area by unravelling the relationships leading to accurate rainfall predictions.

## 1.1 Motivation

Accurate relationships of precipitation to prognostic variables in the parameterization schemes of dynamical weather prediction systems decides the exactness of forecasted rainfall. The linkage becomes even important in the context of monsoons, which are considered the bread and butter of a large section of agrarian economies dependent on rainfall as a source of water supply. For the Indian region alone, the monsoon precipitation's seasonal changes are crucial for the two billion-plus inhabitants. June to September is considered the country's summer monsoon period, as most rainfall occurs during this interval. For the south-eastern region, this period is from October through November. With a standard deviation of sub 10%, the interannual variability as a summation during the monsoon period plays a significant role in deciding the country's agricultural output fate [3]. Error minimization in the model precipitation is also important for the hydroclimate of other regions such as North America which is ravaged by typhoons.

Precipitation over the global land regions has been shown to have global teleconnections with large scale forcings such as El Nino Southern Oscillation and North Atlantic Oscillation. These relationships are an active area of research and various studies focusing on regional and global precipitation delve into the question of strengthening and weaking relationships. For example, [4] argued that the relationship between ENSO and Indian monsoon is breaking down due to global warming. [5] and [6] however subsequently argued that the ENSO-Indian monsoon relationship occurs in epochs. Hence the researchers working on the precipitation and monsoons try to unravel the physical mechanisms behind the nonlinear complex climate system. The studies on precipitation have traditionally been performed using numerical models of the weather and climate [7], which solve partial differential equations of the atmosphere-ocean-land coupled systems in addition to the observational datasets collected from field surveys.

The enormous socio-economic benefits of improved precipitation forecasts have driven dedicated efforts towards enhanced accuracy of the predictions (for example, the Monsoon Mission in India [8,9]). These efforts have shown tremendous economic benefits (e.g., NCAER report 2020 [10]). However, the nonlinear behaviour of the real weather has led to missed forecasts of high-impact extreme events and natural disasters, such as the missed forecast of Kerala (South India) floods in 2018 and the Indian summer monsoon drought in 2002. The missed predictions have primarily been due to the missing physics and parameterizations in the weather and climate models. Hence, the improvements in the physical parameterizations of atmosphere-ocean-land coupled models is the way forward towards improved precipitation forecasts.

### 1.2 Deep learning for weather and climate science

In the last decade, deep learning has emerged as a potential methodology to solve complex, nonlinear problems by unwrapping the nonlinearities in different layers of the deep neural network ([11]). The developments, particularly in computer vision, have led to accurate solutions of problems such as recognizing handwritten digits and others that were not possible until ten years ago. These developments have come about due to the availability of hardware capable of performing memory-intensive convolution operations, which was not possible when convolution neural networks were first proposed [12]. Moreover, software stack development such as open-source libraries (TensorFlow, PyTorch, Theano and others) has led to lower the barriers. The nonlinear operators that have gained prominence in the computer vision community can be directly applied to weather and climate science problems, particularly the problem of deciphering accurate precipitation forecasts in the numerical weather prediction models. It has to be remembered that the advancements that have come about on the shoulders of the dynamical models are not to be thrown away for deep learning, instead they should be complemented by this methodology. Hybrid models that take in the input as fields that are better represented in the models and output the variables that have biases can help improve the skill of error-prone parameters in the dynamical models. [13] note that deep learning can be applied to geoscience problems and with the abundant growth of data and computing resources, machine learning advancements have yielded transformative results across various scientific disciplines. They also point to machine learning becoming a popular approach to the problems of transformation and anomaly detection and geoscientific classification ([13]). The use of neural networks to detect extreme weather patterns replacing traditional threshold-based analysis [13] is one such example. Classical machine-learning approaches (random forests, feed-forward networks, kernel methods) are frequently subject to handcrafted data features [13]. Deep learning is a data-hungry method that can learn the complex mapping between inputs and outputs. Deep learning, which has shown excellent results in a wide variety of fields, can also be used to forecast precipitation [14]. The availability of high-performance computing systems has made it possible to use large datasets to train deep learning models. Understanding the problems of turbulence and nonlinearities as bottlenecks for weather and climate models, deep learning offers immense hope in untangling the near-exact solutions of precipitation forecasting in near future. In [15], author lays down a detailed survey of the potential problems in climate change that can be solved by machine learning. Further, [16] have shown the ability of deep learning methods such as reservoir computing to predict the chaotic Lorenz system. In a recent study by [17], physics-inspired data science has been shown to lead to improved predictions of the turbulent flows. In [18], a feedforward neural network for emulating super parameterization in a climate model. However, convolutional neural networks have been shown to perform better than feedforward neural networks in the computer vision research. The weather systems in the tropics are also turbulent and this is an area where the models generally show poor results.

### 1.3 Related work and critical analysis

In recent times, studies have attempted to use deep learning for weather prediction. In [19], deep convolutional neural networks (DLWP) applied deep convolution neural networks known as UNET is applied to the 500-hPa geopotential height reanalysis dataset. In an improved version, [20] (DLWP-CS) transformed the spherical global data to the cubed sphere and showed improved predictions. The minimization of spherical distortion by the six-faced of the cubed sphere, which could now be taken as images, helped the model to reduce its errors. This was required as most of the developments in deep learning have taken place in the field of computer vision, which involves images on which various mathematical operations such as convolution, nonlinear activation, max-pooling and others are applied. The deep learning libraries that take the advantage of GPU-based CUDA programming assume an image as a default input dataset. [20] compared their results with the numerical weather prediction (NWP) models and found that the cubed sphere model outperformed the IFS42, the T42 spectral resolution model used at the European Centre for Medium-Range Weather Forecasting.

***Table 1:*** *Comparison of machine learning / deep learning studies focused on improved precipitation forecasts*

| Study | A | B | C | D | E | F | G | H | I |
|---|---|---|---|---|---|---|---|---|---|
| DLWP [19] | ✓ | ✓ | x | x | x | ✓ | x | ✓ | x |
| DLWP-CS [20] | ✓ | ✓ | x | x | x | ✓ | x | ✓ | x |
| ResCu [21] | ✓ | ✓ | x | x | ✓ | x | ✓ | ✓ | ✓ |
| NN parameterization [22] | ✓ | ✓ | x | x | ✓ | x | ✓ | x | x |
| ML Atmospheric Forecast Model [23] | ✓ | ✓ | x | x | ✓ | ✓ | x | x | x |
| GAN weather forecast [24] | x | x | x | ✓ | x | x | x | ✓ | x |
| Physics-informed NN [25] | ✓ | ✓ | x | x | ✓ | x | ✓ | x | x |
| NN postprocessing [26] | x | x | x | ✓ | ✓ | ✓ | ✓ | x | x |
| Deep learning GCM [27] | ✓ | ✓ | x | x | x | x | x | ✓ | x |
| Deep learning different complexity GCM [28] | ✓ | ✓ | x | x | x | x | ✓ | ✓ | x |
| RF parameterization [29] | ✓ | ✓ | x | ✓ | ✓ | x | ✓ | x | x |
| Deep learning cloud super-parameterization [30] | ✓ | ✓ | x | ✓ | ✓ | x | ✓ | x | ✓ |
| Deep learning sub grid scale processes [31] | ✓ | ✓ | x | ✓ | ✓ | x | ✓ | x | ✓ |
| Modified DLWP-CS (This study) | ✓ | ✓ | ✓ | ✓ | ✓ | ✓ | ✓ | ✓ | ✓ |

*Abbreviations: **A:** Global region of interest/global datasets, **B**: Technique requires operations over the whole globe, simulate patterns of global teleconnections, **C**: Considers minimizing or removing spherical distortion error due to spherical nature of global datasets of Earth system sciences., **D**: Precipitation as target variable, **E**: Aimed at building hybrid dynamical and data-driven models, F: Comparison with real operational system, **G**: Optimized computational cost for use in long-lead numerical weather prediction systems involving multiple variables, **H**: Learning spatial patterns using convolutional neural networks, **I**: Deep-learning model as a parameterization scheme*

The DLWP and DLWP-CS studies use geopotential at time t as input and at t+1 as output. They perform a temporal learning to demonstrate that deep learning can be used for weather prediction. However, for their implementation, the same variables should be present in the input and output. There are several limitations with the DLWP-CS approach which is modified in this study. The shortcomings are as follows:

1.     Computationally challenging specially for the target diagnostic variables requiring multiple inputs as predictors¬¬.
2.     Deterioration of accuracy with increasing lead time.
3.     It is unclear on how to take advantage of the existing skills of state-of-the-art dynamical models.
4.      Suitable as a complete replacement of dynamical models, however with added computational cost and no value addition from existing numerical weather prediction systems.
5.     They use a framework to forecast fields at t+1 time step using the data at time t and develop a time stepping deep learning model. This has been shown as a proof of concept in their study using geopotential at 500 and comparing with forecast from the TIGGE models. Precipitation is more complex than the geopotential used in their study.
6.     DLWP-CS is limited in its capacity for real applications such as forecasting of precipitation, soil moisture or other diagnostic fields. These variables are dependent on many prognostic variables. For using DLWP-CS for such real-world forecasts iteratively, forecasts of all the variables would have to be required which (i) is highly compute expensive, (ii) would not use the advances in the traditional dynamical models such as the realistic accuracies in fields such as winds and temperature in the traditional models.

In addition to DLWP and DLWP-CS ([19,20]), various other studies have tried to propose solutions to various aspects of weather forecasting problem. Table 1 shows a comparison of the state-of-the-art studies and our implementation. The limitations of some of the previous studies (Table 1) are as follows:
ResCu proposed [21] uses a residual UNET to learn the outputs of super-parameterized Community Atmosphere Model (CAM). They however do not use the cubed-sphere transformation to minimize the spherical distortion error, learn the model world and not real observational or reanalysis world, and do not compare their model outputs with any baseline operational system.
2. NN parameterization [22] learns the datasets generated by an atmospheric model using three-layer neural networks. They, however, do not consider deep-learning, spherical distortion, spatial patterns and comparison with operational forecasts.
3. Machine Learning based atmospheric forecast model [23] showed that reservoir computing based model can outperform persistence and climatology. However, it requires numerous additions such as adjustment for spherical distortion, comparison with operational system, learning spatial patterns and others as shown in Table 1.
4. Generative Adversarial Network (GAN) based weather forecasting is employed by [24] on reanalysis dataset over Europe. Generative models, though, being an impressive methodology, the study uses regional dataset over a small period of four years. Since the atmosphere over the planet Earth is contiguous with global teleconnections changing patterns in different regions, this study needs the use of global datasets, and also addressing the spherical distortion errors.
5. [25] use physics-informed deep neural networks to learn the output of shallow water equation model. This is a theoretical work aimed at improved weather prediction. However, it requires various add-ons to be near to the real system which can lead to the advancements.

NN post-processing [26] shows improved performance by neural networks based postprocessing applied to weather forecast outputs. The study is focused over Germany and compares the results with operational postprocessing techniques. It can be scaled up to a global domain for improving the precipitation outputs in global weather and climate models.

7. [27] use deep convolutional neural networks to learn the outputs of a general circulation model. This work needs to be built upon to learn real world datasets and also account for spherical distortion. Similar improvements need to be added to [28]

8. Random Forest (RF) based parameterization was applied by [29] to show that global high resolution model output can be used to train parameterizations using machine learning.

### 1.4 Our Contributions

The aim of this work is to serve as a proof of concept for the development of a system to assist the dynamical models improve their predictions wherein further work will use the forecasted variables such as temperature, pressure and use them as predictors to target precipitation. Our aim is not to remove the dynamical model completely and do a full-fledged prediction on its own. We assume perfect conditions of meteorological fields (ERA5 weatherbench reanalysis) in the deep learning-based model as input to estimate the reanalysis precipitation, which are similar to the prognostic fields, saved at the first forecast, and used to generate precipitation in the operational dynamical models.

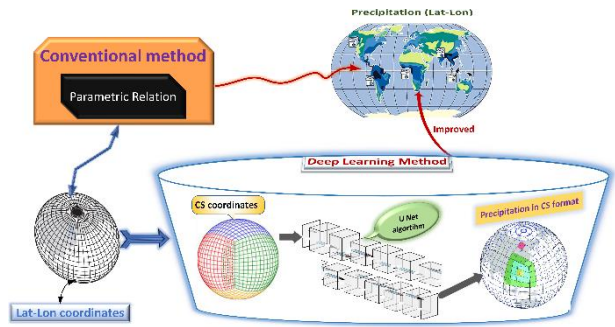

*Figure 1*: Schematic showing the UNET modified DLWP-CS based deep-learning algorithm for learning Earth-system physics towards improved precipitation forecasts

We modify DLWP-CS to decipher the physical relationships leading to precipitation forecasts in the dynamical models. We modify the architecture of cubed-sphere CNN used in DLWP-CS to learn the function from one or many variables to target precipitation at the same time step. Rather than constructing the deep learning architecture to perform time-stepped predictions as in DLWP-CS, i.e., with the input at t, prediction at t+1, we perform a mapping from different variables to precipitation. This technique is expected to eventually lead to improvements in the dynamical models to map rainfall from other fields predicted by the models with reasonable accuracy. To map precipitation from other fields, DLWP-CS would have to be trained with all the variables as input and outputs to perform an iterative forecast, which would be very expensive computationally.

*Table 2*: Details of the convolutional neural network architecture of residual UNET applied on cubed-sphere inputs

| Layer | Filters | Filter size | Output shape | Trainable params |
|---|---|---|---|---|
| Input | | | (6, 96, 96, 1) | |
| Conv2D–CubeSphere | 32 | 3 × 3 | (6, 96, 96, 32) | 6,976 |
| Conv2D–CubeSphere (1) | 32 | 3 × 3 | (6, 96, 96, 32) | 18,496 |
| AveragePooling2D | | 2 × 2 | (6, 48, 48, 32) | |
| Conv2D–CubeSphere | 64 | 3 × 3 | (6, 48, 48, 64) | 36,992 |
| Conv2D–CubeSphere (2) | 64 | 3 × 3 | (6, 48, 48, 64) | 73,856 |
| AveragePooling2D | | 2 × 2 | (6, 24, 24, 64) | |
| Conv2D–CubeSphere | 128 | 3 × 3 | (6, 24, 24, 128) | 147,712 |
| Conv2D–CubeSphere | 64 | 3 × 3 | (6, 24, 24, 64) | 147,584 |
| UpSampling2D | | 2 × 2 | (6, 48, 48, 64) | |
| Concatenate (2) | | | (6, 48, 48, 128) | |
| Conv2D–CubeSphere | 64 | 3 × 3 | (6, 48, 48, 64) | 147,584 |
| Conv2D–CubeSphere | 32 | 3 × 3 | (6, 48, 48, 32) | 36,928 |
| UpSampling2D | | 2 × 2 | (6, 96, 96, 32) | |
| Concatenate (1) | | | (6, 96, 96, 64) | |
| Conv2D–CubeSphere | 32 | 3 × 3 | (6, 96, 96, 32) | 36,928 |
| Conv2D–CubeSphere | 1 | 1 × 1 | (6, 96, 96, 1) | 528 |

Our implementation is similar to the parameterization schemes as they exist in the present form in numerical weather prediction systems. It is the present erroneous parameterization schemes in dynamical models that are expected to improve from our work and the studies to follow. Our implementation offers the following benefits as compared to the literature:
1.      Computationally efficient
2.      Use variables from dynamical models with reasonable accuracy as predictors
3.      Can be used to develop hybrid models
4.      Global datasets are used in addition to convolutions operated on the global data. This would allow for the realistic global teleconnections to be represented by the deep learning model. In contrast to the studies such as [24], which is unable to achieve accuracies in precipitation by using limited domain, our study is able to surpass the operational dynamical system baseline when precipitation skill is compared.
5.      Spherical distortion is minimized when using two-dimensional image-based computer vision algorithms by the use of cubed sphere mapping.
6.      The model is trained on hourly reanalysis datasets. Hence the relationship developed with precipitation are physically robust.
7.      Previous studies ([21], [22]) have shown that neural networks can simulate the complex models such as super-parameterization Community Atmosphere Model (SPCAM) reasonably. We demonstrate that deep neural networks can simulate the real world as well as a proof of concept using ERA5 data.

8.    Observations to parameterizations: Parameterizations in the numerical weather prediction systems are essential to estimate the diagnostic precipitation and other fields which are not directly computed by the partial differential equations. Traditionally, these schemes are empirically designed by means of fields experiments and then inserting them into the global circulation models to improve the precipitation forecasts. For example, [2] used the observed satellite datasets to improve the radiation fluxes in the model CFSv2 and were able to achieve improved estimates of precipitation. In this study, we propose replacing the parameterization scheme with a deep learning model. The traditional schemes were developed using observational field data over specific regions and not validated for regions in poor and developing economies, and hence have inherent biases in the relationships mapped by them.

## 2. Data and methodology

One problem in applying deep learning to weather prediction problems is the availability of data, baselines and other resources at a place. This problem was addressed in [32] by developing Weatherbench, which provides baselines and dataset based on ERA5 reanalysis to enable applications of data-science on weather forecasting problems. We use Weatherbench data to derive improved precipitation (diagnostic) relationships to the other prognostic variables such as total cloud cover, two meter surface air temperature and total incident solar radiation at the surface. Weatherbench dataset which is basically ERA5 reanalyis data for 11 variables [3d(geopotential, temperature, u_wind, v_wind, specific humidity, relative humidity, vorticity, potential_vorticity) and 2d(toa_incident_solar_radiation, total_cloud_cover, total_precipitation)] is available in three spatial resolutions (1.4 degree, 2.8 degree and 5.6 degree). The 1.4-degree dataset is being used for this work. Owing to the computational and disk-storage constraints, we employ only the 2d variables (total cloud cover, two-meter air temperature and total incident solar radiation) as input, and precipitation as the deep learning model's output. All the fields are first normalized using min-max scaling and preprocessed to a cubed sphere (CS) mapping which is then used to train the model. The cubed-sphere mapping minimizes distortion due to the spherical nature of the data. When the spherical dataset is transformed into CS, to reduce spherical distortions, we need to select the spatial resolution of CS faces. We did a forward and reverse mapping to choose the CS resolution, which minimized the transformation error. We find that for our spatial resolution (1.4 degrees) 1024x1024 would be ideal, however, with the hardware (Intel(R) Xeon(R) CPU E5-2695 v4 @ 2.10GHz) used for the CS mapping, it was possible only to transform one-time step (one hour) at a time to 1024x1024. Hence, we chose 512x512 as the CS resolution to transform 24-time steps (i.e. one day) to CS in a single step. However, after the UNET model was created, it was realized that the NVIDIA Tesla P100 (12 GB) GPU only supported a single batch and single-channel as the input and output i.e. (1,6,512,512,1) to (1,6,512,512,1) where the indices correspond to batches, faces, height, width and channels. Our model intends to map the precipitation from different meteorological fields that would be used as the input channels to the deep learning model. Increasing the batch size or channels led to a memory error. Hence, we further reduced the CS resolution to 96x96 to accommodate more variables and the batch size for smooth training.

For the CS mapping, the map files are generated and reused to save the computation time of the transformation. Since the transformation is an expensive computation, the data which originally comes as yearly files from Weatherbench, is broken into daily files and then the mapping is performed. At the time of training with a single variable, training of three years data is done in a go from 1980 to 2009. The dataset corresponding to 2010 and 2011 are used as validation data while training and the years 2012 to 2015 are using as test set for evaluation. CS mapping is a highly data-intensive task as a 1-hourly dataset at a 1.4-degree spatial resolution is first preprocessed to cubed sphere mapping corresponding to each face resolution of 96x96. This data is then being fed into the model. Dask based parallel programming is used for the input/output (I/O) and other operations. Keras library is used for building the deep learning-based model. A schematic of the CNN that is being used for the ongoing training is shown in Figure 1 and the architecture can be noted from Table 2. NVIDIA Tesla P100 (12 GB RAM) GPU as available on the XC50 machine at Pratyush HPC has been used for this work.

---

**Algorithm 1** Modified DLWP-CS

---

1.  **Normalize** training data (min-max scaling)
2.  **Transform** time x lat x lon input to cubed sphere    mapping time x 6 x h x w
3.  **Train** data generator predictors/inputs: total cloud cover, two metre temperature, solar radiation, target/predictand: precipitation
4.  **for** year = 1980 to 2008
5.  **Train** cubed-sphere inputs, cubed-sphere precipitation targets. fit (Xtrain, ytrain), Xtrain, ytrain at same times.
6.  **Stop** training using early stopping, validation years =   2009,2011
7.  **Save** trained model
8.  **Normalize** test data
9.  **Transform** test data to cubed sphere mapping
10. **for** year = 2012 to 2015
11. **Predict** output using trained model
12. **Compare** trained predictions with GFS daily model output

---

### 2.1 Training

It takes around 17-20 hours for training one epoch which mainly includes 3288 samples of the hourly input data from the three years of data used for training in a single pass with a batch size of 8. Iterative training is performed so that three years of training data is loaded on the data generator in a single pass. It takes around 22 seconds for one step of the epoch to train with the batch size as 8. The number of samples is computed by dividing the number of time steps by the batch size. So, if the three years of data have 26280(8760*3) time steps, the number of samples per epoch is computed as 3285 (26280/8). A learning rate of 1e-4 is used and rectified linear unit, also known as RELU, is used as the nonlinear activation function.

An early stopping algorithm within Keras is used with min_delta as 0, patience as 2, verbose as one and mode as 'auto'. The loss function is a mean squared error, and early stopping is done based on the monitored validation loss. In general, the default number of epochs used is 10000; however, due to early stopping, those numbers of epochs were never realised. The model weights and forecasted precipitation for the 3 models are saved as a separate output to perform testing and future forecasts. The complete training with early stopping took around 40-50 days on an NVIDIA Tesla P100 GPU.

## 2.2 Testing

The global forecasts are generated in a couple of hours from the trained model for the entire period of four years of testing data. This is another advantage of using a deep learning-based system relative to the dynamical numerical weather prediction models that take much longer for similar forecasts. The model's testing is performed for the period 2012-2015 and is compared with the operational IITM-GFS outputs from the T574 model. The deep learning based model uses features such as early stopping, custom CubedSphereConv2D class to map one variable to another and a custom data generator to facilitate training, validation, and testing. In summary, the AI/ML-based system can be considered a sophisticated data-driven parameterisation scheme that is hoped to lead to improved accuracy in the predictions of the Indian monsoon.

## 3. Results and discussion

The evaluation of the UNET based deep learning models is performed on the data for the June-September period of the years 2012-2015 as the operational baseline is available for this duration. The test data predictions for these periods are compared with the corresponding output from the operational model GFS T574 ([33]). The dynamical model inputs initial conditions which are the state of the real Earth system to compute the prognostic fields which are then fed to the parameterization scheme and generate diagnostic fields such as precipitation.

*Table 3*: Details of various deep learning models used in the study

| MODEL | INPUT VARIABLE | OUTPUT VARIABLE | DATA SOURCE | NVAR |
|-------|----------------|-----------------|-------------|------|
| DL1 | Total cloud cover | Precipitation | ERA5 | 1 |
| DL2 | Two-meter air temp | Precipitation | ERA5 | 1 |
| DL3 | Total incident solar radiation | Precipitation | ERA5 | 1 |

Our deep learning-based model works on a similar principle to input the total cloud cover, two metre temperature and total incident solar radiation as individual inputs and output the precipitation as shown in Table 3. Since the GFS model outputs are available at daily frequencies, the output from the deep learning model is transformed to a daily summation from the hourly time scales test predictions. It can be observed from Fig. 2 that there are substantial improvements in the predicted precipitation for DL1 and DL2 relative to the dynamical model, whereas DL3 performed worse than the operational model forecasts. The best model is DL1 which is built using two-meter surface air temperature as an input variable (Table 4).

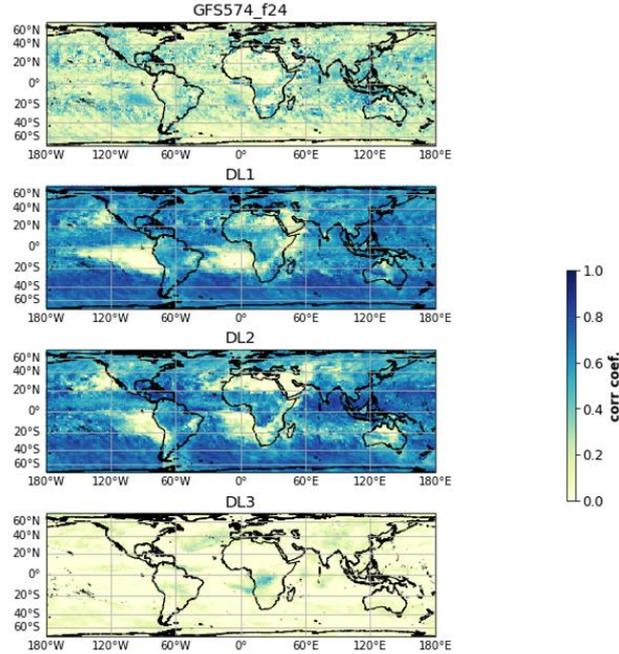

**Figure 2**: Correlation coefficients between GFS (baseline), DL1, DL2, DL3 at each grid point and the ERA5 (Weatherbench) precipitation for the test period 2012-2015

It can be noted that the deep learning-based parameterization is able to model the precipitation reasonably well over the land regions relative to the oceans (Fig. 2). In particular, the rainfall dependent monsoon regions show a correlation coefficient of ~0.6 relative to the value of ~0.3 in the GFS dynamical model baseline (Table 4). It can be seen from the figures that the deep learning model with just a single input as the two-metre surface air temperature and precipitation as a target performs reasonably well relative to the operational dynamical system.

### 4. Conclusions and future work

In this study we modify the DLWP-CS to develop relationships of three variables viz, total cloud cover, two metre temperature and total incident solar radiation to precipitation in the real-world ERA-5 reanalysis product. DLWP-CS proposes the use of a residual UNET based deep learning model to learn temporal relationships between same / different variables and improves on the previous studies by incorporating cubed sphere mapping to reduce errors due to spherical nature of the datasets. However, for their model, same number of inputs and outputs would be required. For variables such as precipitation which are dependent on other fields of the atmosphere, ocean and land, DLWP-CS would be computationally expensive. Moreover, their deep learning model can drift at large lead times taking no advantage of existing state of the art existing dynamical model. Moreover, it has also not been tested yet for the full-fledged precipitation forecasts, so it is uncertain whether the fields such as temperature and winds which are used as precursors for precipitation, would be reasonably accurate using DLWP-CS. We discuss the limitations of their method and the advantages in learning spatial features to target precipitation rather than temporal learning. The benefits of our modified DLWP-CS approach are towards building hybrid data-driven and dynamical models, and using optimum amount of computing resources.

***Table 4***. Correlation coefficients over different land-regions from GFS, DL1, DL2 and DL3 models and ERA5 precipitation for the test years 2012-2015

| REGION | GFS (BASELINE) | DL1 (TCC) | DL2 (TAS) | DL3 (TISR) |
|---|---|---|---|---|
| CANADA (50-70N, 130-90W) | 0.196 | 0.627 | 0.492 | 0.020 |
| NORTH ASIA (50-70N, 25-140E) | 0.237 | 0.635 | 0.541 | 0.033 |
| EUROPE (44-54N, 40-100E) | 0.276 | 0.639 | 0.549 | 0.006 |
| UNITED STATES (32-50N, 110-80W) | 0.210 | 0.622 | 0.501 | 0.015 |
| CENTRAL ASIA (35-50N, 40-100E) | 0.234 | 0.526 | 0.375 | 0.072 |
| AMAZON (5N-10S, 50-70W) | 0.164 | 0.497 | 0.584 | -0.09 |
| EQUATORIAL AFRICA (10N-10S, 14-35E) | 0.211 | 0.470 | 0.594 | 0.194 |
| INDIA (18-35N, 70-90E) | 0.316 | 0.619 | 0.587 | 0.073 |

We show that a residual UNET based architecture can be used to learn earth system physics from the global four-dimensional sptaio-temporal datasets of the Earth system. In particular, the problem of precipitation formation inside the numerical weather prediction systems is addressed. We use three deep learning-based models with the same architecture (UNET) but with different input variables: total cloud cover, two-meter surface air temperature, and total incident solar radiation.

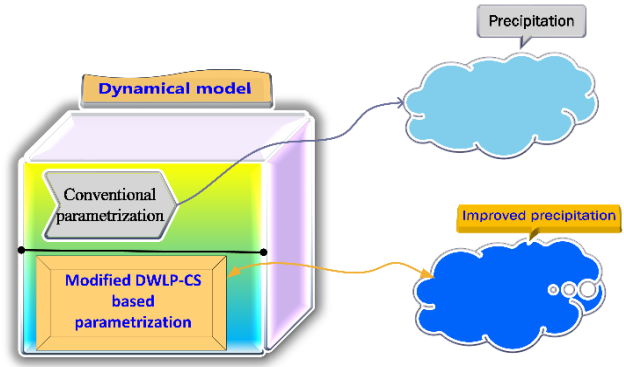

**Figure 3**: Proposed framework for development of improved hybrid numerical weather prediction systems using modified DLWP-CS

This study used just a single variable as input due to computational and time constraints. In future, adding more variables in addition to the hyperparameter tuning is expected to lead to further improvements. Also, the input dataset that has been used is from the reanalysis products. In a more real-world scenario, those inputs may be biased and will propagate the errors in the neural network. The comparison with the GFS model in this work is only to show that the correlations are comparable to a dynamical model, although, for a pure one-to-one comparison, the model data distribution needs to be provided as an input and the ground truth precipitation as the target variable. We base our work on improving the existing dynamical models and hence target the process by which rainfall is formed in the numerical weather prediction systems. This would require improved parameterizations of the rain in the dynamical models and, hence, more realistic precipitation relationships to other variables. This is done by framing the problem from a temporal mapping to a spatial mapping with the knowledge that the existing models perform relatively well on predicting variables other than precipitation. Hence, our system can build into a complete online hybrid framework that can improve weather forecasts and climate predictions.

### Acknowledgements

The authors acknowledge the use of high performance computational resources at IITM, particularly the NVIDIA Tesla P100 GPU without which this work would not have been possible.

### References

[1] Schiermeier, Q. (2010). The real holes in climate science. Nature News, 463(7279), 284-287.

[2] Goswami, T., Mukhopadhyay, P., Ganai, M., Krishna, R.P.M., Mahakur, M. and Han, J.Y., 2020. How changing cloud water to rain conversion profile impacts on radiation and its linkage to a better Indian summer monsoon rainfall simulation. Theoretical and Applied Climatology, 141, pp.947-958.

[3] Gadgil, S., 2003. The Indian monsoon and its variability. Annual Review of Earth and Planetary Sciences, 31(1), pp.429-467.

[4] K. K. Kumar, B. Rajagopalan, M. A. Cane, On the weakening relationship between the Indian monsoon and ENSO. Science 284, 2156–2159 (1999).

[5] D. Maraun, J. Kurths, Epochs of phase coherence between El Niño/Southern Oscillation and Indian monsoon. Geophy. Res. Lett. 32, L15709 (2005).

[6] Singh, M., Krishnan, R., Goswami, B., Choudhury, A.D., Swapna, P., Vellore, R., Prajeesh, A.G., Sandeep, N., Venkataraman, C., Donner, R.V. and Marwan, N., 2020. Fingerprint of volcanic forcing on the ENSO–Indian monsoon coupling. Science advances, 6(38), p.eaba8164.

[7] Krishnan, R., Singh, M., Vellore, R. and Mujumdar, M., 2020. Progress and Prospects in Weather and Climate Modelling. arXiv preprint arXiv:2011.11353

[8] Rao, S.A., Goswami, B.N., Sahai, A.K., Rajagopal, E.N., Mukhopadhyay, P., Rajeevan, M., Nayak, S., Rathore, L.S., Shenoi, S.S.C., Ramesh, K.J. and Nanjundiah, R.S., 2019. Monsoon Mission: A targeted activity to improve monsoon

[9] prediction across scales. Bulletin of the American Meteorological Society, 100(12), pp.2509-2532.

[10] Rajeevan, M.N. and Santos, J., India's Monsoon Mission., 2020

[11] NCAER, 2020: Estimating the economic benefits of Investment in Monsoon Mission and High Performance Computing facilities, National Council of Applied Economic Research, New Delhi, India, Report No. 20200701, 76pp, https://moes.gov.in/writereaddata/files/Economic_Benefits_NCAER_Report.pdf

[12] Zeiler, M.D. and Fergus, R., 2014, September. Visualising and understanding convolutional networks. In European conference on computer vision (pp. 818-833). Springer, Cham.

[13] LeCun, Y., Bottou, L., Bengio, Y. and Haffner, P., 1998. Gradient-based learning applied to document recognition. Proceedings of the IEEE, 86(11), pp.2278-2324.

[14] Reichstein, M., Camps-Valls, G., Stevens, B., Jung, M., Denzler, J. and Carvalhais, N., 2019. Deep learning and process understanding for data-driven Earth system science. Nature, 566(7743), pp.195-204.

[15] Shi, X., Gao, Z., Lausen, L., Wang, H., Yeung, D.Y., Wong, W.K. and Woo, W.C., 2017. Deep learning for precipitation nowcasting: A benchmark and a new model. In Advances in neural information processing systems (pp. 5617-5627).

[16] Rolnick, D., Donti, P.L., Kaack, L.H., Kochanski, K., Lacoste, A., Sankaran, K., Ross, A.S., Milojevic-Dupont, N., Jaques, N., Waldman-Brown, A. and Luccioni, A., 2019. Tackling climate change with machine learning. arXiv preprint arXiv:1906.05433.

[17] Chattopadhyay, A., Hassanzadeh, P. and Subramanian, D., 2020. Data-driven predictions of a multiscale Lorenz 96 chaotic system using machine-learning methods: reservoir computing, artificial neural network, and long short-term memory network. Nonlinear Processes in Geophysics, 27(3), pp.373-389.

[18] Wang, R., Kashinath, K., Mustafa, M., Albert, A. and Yu, R., 2020, August. Towards physics-informed deep learning for turbulent flow prediction. In Proceedings of the 26th ACM SIGKDD International Conference on Knowledge Discovery & Data Mining (pp. 1457-1466).

[19] Mooers, G., Pritchard, M., Beucler, T., Ott, J., Yacalis, G., Baldi, P. and Gentine, P., 2020. Assessing the Potential of Deep Learning for Emulating Cloud Superparameterization in Climate Models with Real-Geography Boundary Conditions. arXiv preprint arXiv:2010.12996.

[20] Weyn, J.A., Durran, D.R. and Caruana, R., 2019. Can machines learn to predict weather? Using deep learning to predict gridded 500‐hPa geopotential height from historical weather data. Journal of Advances in Modeling Earth Systems, 11(8), pp.2680-2693.

[21] Weyn, J.A., Durran, D.R. and Caruana, R., 2020. Improving data-driven global weather prediction using deep convolutional neural networks on a cubed sphere. arXiv preprint arXiv:2003.11927

[22] Han, Y., Zhang, G.J., Huang, X. and Wang, Y., 2020. A moist physics parameterization based on deep learning. Journal of Advances in Modeling Earth Systems, 12(9), pp.e2020MS002076.

[23] Yuval, J., O'Gorman, P.A. and Hill, C.N., 2021. Use of neural networks for stable, accurate and physically consistent parameterization of subgrid atmospheric processes with good performance at reduced precision. Geophysical Research Letters, 48(6), p.e2020GL091363.

[24] Arcomano, T., Szunyogh, I., Pathak, J., Wikner, A., Hunt, B.R. and Ott, E., 2020. A Machine Learning‐Based Global Atmospheric Forecast Model. Geophysical Research Letters, 47(9), p.e2020GL087776.

[25] Bihlo, A., 2021. A generative adversarial network approach to (ensemble) weather prediction. Neural Networks, 139, pp.1-16.

[26] Bihlo, A. and Popovych, R.O., 2021. Physics-informed neural networks for the shallow-water equations on the sphere. arXiv preprint arXiv:2104.00615.

[27] Rasp, S. and Lerch, S., 2018. Neural networks for postprocessing ensemble weather forecasts. Monthly Weather Review, 146(11), pp.3885-3900.

[28] Scher, S., 2018. Toward data‐driven weather and climate forecasting: Approximating a simple general circulation model with deep learning. Geophysical Research Letters, 45(22), pp.12-616.

[29] Scher, S. and Messori, G., 2019. Weather and climate forecasting with neural networks: using general circulation models (GCMs) with different complexity as a study ground. Geoscientific Model Development, 12(7), pp.2797-2809.

[30] Yuval, J. and O'Gorman, P.A., 2020. Stable machine-learning parameterization of subgrid processes for climate modeling at a range of resolutions. Nature communications, 11(1), pp.1-10.

[31] Mooers, G., Pritchard, M., Beucler, T., Ott, J., Yacalis, G., Baldi, P. and Gentine, P., 2020. Assessing the Potential of Deep Learning for Emulating Cloud Superparameterization in Climate Models with Real-Geography Boundary Conditions. arXiv preprint arXiv:2010.12996

[32] Rasp, S., Pritchard, M.S. and Gentine, P., 2018. Deep learning to represent subgrid processes in climate models. Proceedings of the National Academy of Sciences, 115(39), pp.9684-9689.

[33] Rasp, S., Dueben, P.D., Scher, S., Weyn, J.A., Mouatadid, S. and Thuerey, N., 2020. WeatherBench: A benchmark dataset for data-driven weather forecasting. arXiv preprint arXiv:2002.00469.

[34] Prasad V S, Mohandas S, Das Gupta M, Rajagopal E N and Datta S K 2011 Implementation of upgraded global forecasting systems (T382L64 and T574L64) at NCMRWF; NCMRWF Technical Report No. NCMR/TR/5/2011 May 2011, 72p, http://www.ncmrwf.gov.in/ncmrwf/gfs.report.final.pdf